Silvia Ferro ✉ ⓘ ; Matteo Faganello ⓘ ; Francesco Califano ⓘ ; Fabio Bacchini ⓘ

# Comparative simulations of Kelvin–Helmholtz induced magnetic reconnection at the Earth's magnetospheric flanks

Silvia Ferro,[1,a)] Matteo Faganello,[2] Francesco Califano,[3] and Fabio Bacchini[1,4]

AFFILIATIONS

[1]Centre for Mathematical Plasma Astrophysics, Department of Mathematics, KU Leuven, Celestijnenlaan 200B, B-3001 Leuven, Belgium
[2]Aix-Marseille University, CNRS, PIIM UMR 7345, Marseille, France
[3]Dipartimento di Fisica "E. Fermi", Università di Pisa, 56127 Pisa, PI, Italy
[4]Royal Belgian Institute for Space Aeronomy, Solar-Terrestrial Centre of Excellence, Ringlaan 3, 1180 Uccle, Belgium

Note: This paper is part of the Special Topic on 1st European Conference on Magnetic Reconnection in Plasmas.
a)Author to whom correspondence should be addressed: silvia.ferro@kuleuven.be

ABSTRACT

This study presents three-dimensional (3D) resistive Hall-magnetohydrodynamic simulations of the Kelvin–Helmholtz instability (KHI) dynamics at Earth's magnetospheric flanks during northward interplanetary magnetic field periods. By comparing two simulations with and without initial magnetic shear, we analyze the impact of distinct magnetic field orientations on plasma dynamics and magnetic reconnection events taking into account 3D mechanisms, such as KHI high latitude stabilization. The identical nature of the simulations, except for the presence/absence of an initial magnetic shear, enables, for the first time, a complete and coherent comparative analysis of the latitudinal distribution of KH vortices, current sheets, reconnection events, and the evolution of the mixing layer. In one configuration, a uniform magnetic field leads to double mid-latitude reconnection (MLR), while in the other, magnetic shear induces both type I vortex-induced reconnection (VIR) and MLR. Notably, the type I VIR observed in this second scenario results from the combined action of line advection and vortex-induced current sheet pinching (the classic mechanism driving two-dimensional type I VIR). Of particular importance is our quantification of newly closed field lines that experienced double reconnection, ultimately becoming embedded in solar wind plasma at low latitudes while remaining connected to magnetospheric plasma at high latitudes. The varying abundance of such lines in the two simulations holds implications for plasma transport at the magnetopause.

## I. INTRODUCTION

Magnetic reconnection is a fundamental mechanism capable of altering the topology of magnetic field lines within plasmas. Specifically, it plays a significant role within the Earth's magnetosphere (MS). Reconnection can transfer the energy stored in the global, large-scale structure of the magnetic field to kinetic scales, and it releases this energy locally while also modifying the line connectivity. The classic magnetospheric example is provided by the Dungey cycle[1] during periods of southward interplanetary magnetic field (IMF): IMF lines and magnetic flux, advected by the solar wind (SW), pile up in front of the magnetopause, where the electric current is strongly enhanced. At scales comparable with the electron inertial length, the frozen-in condition breaks, and reconnection occurs.[2] Thus, magnetic reconnection modifies the topology of the magnetospheric field and that of the SW, creating "open" field lines connected to the Earth at one pole only. This global topological modification of the MS allows for plasma recirculation and plasma transport across the magnetopause. Indeed, open field lines can be advected in the nightside region, accumulating magnetic flux along the magnetotail, where finally magnetic reconnection occurs once again.[3] Furthermore, SW particles can easily stream along open field lines and enter the MS.[4,5]

During northward IMF periods, dayside reconnection is ineffective. On the contrary, cusp reconnection[6] can occur, causing plasma transport at dayside and generating a low-latitude boundary layer.[7]

*In situ* observations have proven that during northward IMF periods, the Kelvin–Helmholtz instability (KHI) is active at the low latitude magnetopause.[8–10] There, the KHI mediates the interaction between the SW and the MS and drives reconnection, boosting plasma transport.[4,11] *In situ* observations also emphasize the role of an enhanced local magnetic shear in boosting magnetic reconnection and plasma transport from magnetosheath to magnetosphere,[12] while simulations based on satellite observations have underlined the critical role of the KHI in driving turbulence in the same region.[13] Although the KHI is predominant during northward IMF orientation, southward IMF configurations also offer a wide range of mechanisms and phenomena that interest the formation and evolution of the KHI. The location where particle mixing occurs inside KH vortices, for example, changes according to the IMF orientation.[14,15] Because *in situ* observations for southward IMF orientations are fewer, a greater number of studies have been devoted to northward IMF orientations, and even within the same global configuration, different magnetic field orientations trigger different mechanisms as the KHI develops.

If the IMF is mostly northward but not strictly aligned with the magnetospheric field, magnetic shear arises. In this configuration, the KHI forces reconnection to occur with a mechanism somewhat similar to dayside reconnection: the velocity field imposed by the growing, ideal KHI accumulates magnetic flux at the pre-existing magnetopause current sheet until reconnection occurs. The pinching of the low-latitude current sheet of the magnetopause is, thus, the peculiar signature of type I vortex induced reconnection (VIR).[16,17] Notably, in the case of three-dimensional (3D) simulations that take into account the high-latitude stabilization of the KHI, the classification of events as "Type I VIR" is an oversimplification. Indeed, during the system evolution, both current pinching and differential (with respect to latitude) advection of magnetic field lines are active simultaneously, with consequences on the distribution of the type I VIR reconnection sites that will be explored in this work.

Conversely, when the magnetospheric field and the IMF are sufficiently aligned at low latitudes, current sheets will not develop in those regions. The KHI will create current sheets elsewhere by folding and/or stretching magnetic field lines. In the oversimplified yet widely considered two-dimensional (2D) picture, KH vortices can fold the in-plane (with respect to the 2D dynamical plane) component of the initial field, thereby leading to the development of current sheets and type II VIR.[18,19] In a more realistic three-dimensional configuration, field lines are tied to high-latitude regions, which are less prone to KH development. As a result, lines are advected differently at low and high latitudes, and current sheets are created far away from the equatorial plane, at mid-latitudes, where finally mid-latitude reconnection (MLR) takes place.[20]

The topological modifications generated in the two scenarios above are different. Starting from a configuration with aligned magnetic fields and without an initial current sheet at low latitudes, we provide a reference case where only MLR proceeds almost simultaneously in both hemispheres (double MLR), leading to the formation of "twice-reconnected lines."[21,22] Double-MLR can explain the increase in specific entropy inside the MS during northward periods.[23] Half of these "twice-reconnected lines" are "newly closed," meaning they are connected to both poles but embedded in SW plasma at low latitudes. In the other scenario, as soon as a magnetic shear exists, type I VIR, instead, creates "open" field lines that connect the two plasma regions.

Both mechanisms have been shown to contribute substantially to the transport properties of the flank magnetopause.[5,20,24] Notably, when high-latitude stabilization is taken into account, type I VIR can occur in conjunction with MLR. In fact, while type I VIR is forced by the pinching of a low-latitude current sheet, where the vortices are well-developed, 3D effects related to the differential advection of field lines at low and high latitudes induce MLR. These complex dynamics have been inferred by Magnetospheric Multiscale (MMS) Mission data[25–27] and confirmed by numerical simulations.[28,29] In this way, a mix of open field lines and newly closed lines can be found at the magnetopause.

Here, with the aid of numerical simulations, we perform a comparative study to explore the two potential 3D evolution scenarios observed during northward IMF periods. The two configurations vary in the orientation of the magnetic field, with one featuring a purely northward magnetic field and the other introducing a magnetic rotation and, consequently, a shear in the initial magnetic field. These two simulations represent prototype configurations for studying MLR alone (run A) or a combination of type I VIR and MLR (run B) in a simplified equilibrium with a symmetric magnetic field. By designing two simulations identical in size, numerical resolution, and physical equilibrium quantities, we allow for the first time for a one-to-one comparison between the KHI development in two 3D configurations that differ only for the presence and absence of the initial magnetic shear. Our focus lies specifically on quantifying the amount of reconnected lines generated in both configurations, determining their implications for plasma transport at the magnetopause, and evaluating the relative proportions of open and newly closed magnetic field lines.

This article is structured as follows: In Sec. II, we introduce the plasma models, starting equilibria, and the numerical method. In Sec. III, we compare simulation results, and we discuss them in Sec. IV, while Sec. V is dedicated to conclusions.

## II. PLASMA MODEL AND SIMULATION SET UP
### A. Resistive Hall-MHD model and numerical code

The evolution of the plasma is described by a resistive Hall-MHD quasi-neutral model,[20] which accounts for different evolution for the electron and ion pressures and diamagnetic effects for the electrons. The model equations read

$$\frac{\partial n}{\partial t} + \nabla \cdot (n\boldsymbol{u}) = 0, \quad (1)$$

$$\frac{\partial (n\boldsymbol{u})}{\partial t} + \nabla \cdot \left[ (n\boldsymbol{u}\boldsymbol{u}) + (\Pi \overleftrightarrow{\boldsymbol{I}} - \boldsymbol{B}\boldsymbol{B}) \right] = \boldsymbol{0}, \quad (2)$$

where $n$ is the bulk plasma number density, $\boldsymbol{u}$ the fluid velocity (neglecting electron inertia), $\boldsymbol{B}$ the magnetic field, and $\Pi = P_{\text{th}} + P_{\text{m}}$ is the total pressure (sum of thermal pressure $P_{\text{th}} = P_i + P_e$ and magnetic pressure $P_{\text{m}} = B^2/2$). The thermal pressures of ions and electrons evolve following an adiabatic closure (for each species $\alpha = i, e$),

$$\frac{\partial (n_\alpha S_\alpha)}{\partial t} + \nabla \cdot (n_\alpha S_\alpha \boldsymbol{u}_\alpha) = 0, \quad S_\alpha = P_{\text{th}} n^{-5/3}. \quad (3)$$

To complete the set of equations, we introduce Faraday's and Ampère's laws to describe the evolution of the magnetic field (neglecting the displacement current, given the non-relativistic regime of the magnetospheric plasma), along with the evolution of the current density $\boldsymbol{j}$,

$$\frac{\partial \boldsymbol{B}}{\partial t} = -\nabla \times \boldsymbol{E}, \quad \boldsymbol{j} = \nabla \times \boldsymbol{B} = n(\boldsymbol{u}_i - \boldsymbol{u}_e). \quad (4)$$

Here, $\boldsymbol{u}_i$ and $\boldsymbol{u}_e$ are the ion and electron fluid velocities, with $\boldsymbol{u}_i \simeq \boldsymbol{u}$ and $\boldsymbol{u}_e \simeq \boldsymbol{u}_i - \boldsymbol{j}/n$ in the limit of massless electrons. All quantities are normalized using the ion mass $m_i$, the ion skin depth $d_i$, and the ion cyclotron frequency $\Omega_c$. Finally, the generalized Ohm's law accounting for the electric field $\boldsymbol{E}$ reads

$$\boldsymbol{E} = -\boldsymbol{u} \times \boldsymbol{B} + \frac{\boldsymbol{j} \times \boldsymbol{B}}{n} - \frac{1}{n}\nabla P_e + \eta \boldsymbol{j}. \quad (5)$$

The introduction of an artificial, small resistivity $\eta \sim 10^{-3}$ in Eq. (5) ultimately enables the occurrence of magnetic reconnection. With our choice of $\eta$, justified by earlier 2D studies,[30,31] the resistive diffusion timescale is $\tau_R \sim L_{eq}^2 \eta^{-1} \approx 10^4$, where $L_{eq} = 3d_i$ is the gradient length of the equilibrium configuration. This timescale is far longer than our simulation time. Consequently, at large scales, the magnetic field is frozen-in in the fluid motion generated by the KH dynamics, and inside the vortices magnetic diffusion is negligible. Furthermore, the Hall term plays a critical role in $\beta \sim 1$ plasmas for describing the decoupling between the magnetic and the ion dynamics at nearly the ion skin depth. Most importantly, taking into account the Hall term enables the transition to fast reconnection characterized by a reconnection rate weakly dependent on the value of the resistivity.[32] Even if type I VIR is a kind of forced reconnection, with a reconnection rate that is more or less insensitive to the resistivity value even in a simplified MHD approach,[16] this is not the case for MLR that proceeds as spontaneous reconnection.[20] Since the conditions for MLR are dynamically created by the large-scale motion, the speed at which reconnection proceeds can be critical for determining the evolution of the magnetic topology.[31] We, thus, argue that including the Hall term in our model provides a more comprehensive treatment of the reconnection dynamics induced by KH vortices.

We use a fourth-order Runge–Kutta scheme for the numerical integration of Eqs. (1)–(5) and calculate spatial derivatives using sixth-order explicit finite differences in the $y$- and $z$-directions. In the $x$-direction, we implement a sixth-order implicit compact scheme with spectral-like resolution,[33] as this is the direction in which smaller scale dynamics emerge. We impose periodic conditions at the $y$- and $z$-boundaries. The $x$-boundaries are open, so we apply conditions based on magnetohydrodynamic characteristics, allowing Alfvénic and magnetosonic perturbations to freely exit the numerical domain.[34,35]

### B. Geometrical configuration and equilibrium

We adopt a 3D slab configuration to schematically model the structure of Earth's magnetopause flank, including 3D mechanisms, such as the stabilization of the KHI at high latitudes. In our frame of reference, the $x$-axis is perpendicular to the undisturbed magnetopause, the $y$-axis aligns with the SW flow direction, and the $z$-axis corresponds to the northward direction. In order to track the evolution of the KH-induced vortices, the numerical box moves[21] at the phase velocity of the KHI along the MS flank toward the tail region, effectively remaining fixed in the rest frame of the vortex.

For the initial configuration, we consider a 2D Grad–Shafranov equilibrium satisfying

$$\nabla^2 \psi = -4\pi \frac{d\Pi}{d\psi}, \quad (6)$$

where $\psi$ is the flux function. The corresponding equilibrium magnetic field reads

$$\boldsymbol{B}_{eq}(x,z) = \nabla \psi \times \hat{y} + B_{y,eq}(\psi)\hat{y}. \quad (7)$$

In Eq. (7), the first term represents the (nearly) northward MS field. The second term $B_{y,eq}$ corresponds to the flow-aligned component of the IMF, which reproduces the overall magnetic field rotation when combined with the northward magnetic field component. All other physical quantities are obtained from the flux function and are by construction constant along magnetic field lines. Setting $\Pi$ constant with respect to $\psi$, the flux function can be written as a harmonic function

$$\psi = \frac{1}{2}\left[\left(1+\frac{1}{\delta}\right)\left(x-\frac{L_x}{2}\right) + \left(1-\frac{1}{\delta}\right)\frac{L_z}{2\pi}\sinh\frac{2\pi(x-L_x/2)}{L_z}\right.$$
$$\left. \times \cos\frac{2\pi(z-L_z/2)}{L_z}\right]. \quad (8)$$

The resulting equilibrium quantities are given by

$$B_{x,eq} = \frac{1}{2}\left(1-\frac{1}{\delta}\right)\sinh\frac{2\pi(x-L_x/2)}{L_z}\sin\frac{2\pi(z-L_z/2)}{L_z},$$
$$B_{y,eq} = \tan\frac{\phi}{2}\tanh\left(\frac{\psi}{L_{eq}}\right),$$
$$B_{z,eq} = \frac{1}{2}\left[\left(1+\frac{1}{\delta}\right) + \left(1-\frac{1}{\delta}\right)\cosh\frac{2\pi(x-L_x/2)}{L_z}\right. \quad (9)$$
$$\left. \times \cos\frac{2\pi(z-L_z/2)}{L_z}\right],$$
$$u_{y,eq} = \frac{\Delta u}{2}\tanh\frac{\psi}{L_{eq}}, \quad \text{and} \quad u_{x,eq} = u_{z,eq} = 0.$$

The equilibrium number density is constant, $n_{eq} = 1$. The initial thermal pressure is set so that $\Pi$ is constant, with $P_{th}|_{\psi=0} = 1$ and $P_i/P_e = 4$, as inspired by earlier works.[29,36]

Furthermore, the chosen value $\delta \ll L_x < L_z/4$ ensures that the dominant component of the magnetic field in the $(x,z)$-plane is aligned with the northward direction. Setting $L_{eq} = 3d_i$ allows the KHI to develop at large, nearly MHD scales, with the wavelength of the fastest growing mode (FGM) measuring approximately $\lambda_{FGM} \sim 47 d_i$. Finally, we set $L_y = 2\lambda_{FGM}$ in order to let multiple vortices develop in the numerical box. Since $L_z > 8\lambda_{FGM}$, stable high-latitude regions are far enough from the unstable equatorial one so that KH vortices can grow and even pair,[37] merging distinct coherent structures into a single one.

In Table I, we list the main parameters of the numerical box and the equilibrium configuration. We have performed two identical runs differing only by the presence (or absence) of shear in the initial magnetic field configuration. This was achieved using the magnetic shear angle $\phi$ [see $B_{y,eq}$ in Eq. (9)], which represents the total angle between the MS and the SW magnetic fields. In Fig. 1, we show in the left panel the results of the first simulation, "run A," in which $\phi_A = 0$. This implies no rotation in the magnetic field, hence, $B_{y,eq} = 0$. The right panel shows the second simulation, "run B," with $\phi_B \simeq 18°$. This introduces a magnetic rotation at equilibrium and, therefore, the presence of an initial magnetic shear in the system. A consequence of this initial setting is that at $t=0$, in run B, there is a small initial current sheet (absent in run A) at the frontier between the two plasma regions.

TABLE I. Summary of the relevant parameters characterizing the numerical boxes and the equilibrium configurations: the box dimensions $(L_x, L_y, L_z)$, the number of grid points $(N_x, N_y, N_z)$, the equilibrium scale length $L_{eq}$, the magnetic shear angle of both runs $\phi_{A,B}$, the velocity jump across the two plasma regions $\Delta u$, and the parameter $\delta$ determining the ratio between $B_z(x=0, z=L_z/2)$ and $B_z(x=0, z=0)$.

| Parameter | Value | Parameter | Value |
| --- | --- | --- | --- |
| $L_x$ | 90 | $N_x$ | 600 |
| $L_y$ | 94 | $N_y$ | 512 |
| $L_z$ | 377 | $N_z$ | 512 |
| $\delta$ | 3 | $L_{eq}$ | 3 |
| $\Delta u$ | 1 | $\phi_{A,B}$ | 0°, 18° |

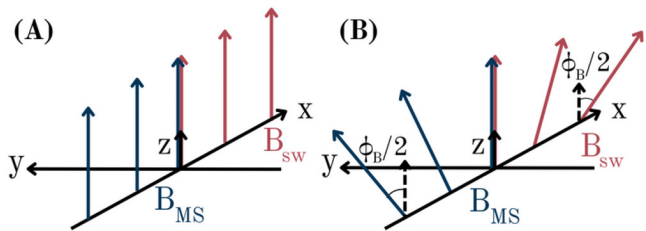

FIG. 1. Sketch of the initial magnetic field configuration. In run A, the magnetic shear angle $\phi_A = 0$; therefore, both the MS and SW magnetic fields are aligned northward. In run B, the presence of $\phi_B \simeq 18°$ introduces a rotation: at the magnetopause (the frontier), the magnetic fields are aligned with the northward direction, but both fields are inclined by $\pm \phi_B/2$ for $x \to 0, L_x$.

The chosen velocity profile [see Eq. (9)] is the same for both runs and has stronger gradients at the equator and weaker at high latitudes. This is shown in Fig. 2, where we plot a slice at $y = 0$ through the numerical box for run A. In particular, we show the hourglass-like configuration of the initial magnetic field lines (which correspond to the isocontours of the magnetic flux) in the $(x, z)$-plane, with lines getting closer together at the equatorial plane $(z \sim L_z/2)$ than at high latitudes $(z \sim 0$ and $z \sim L_z)$. Since the growth rate of the instability $\gamma_{KH}$ is proportional to the velocity gradient,[38] this equilibrium drives the development of the KHI far earlier in the equatorial plane, while high latitude regions remain "quiet" throughout the simulation. This configuration effectively captures the MS flank dynamics. The choice of a uniform density enables us to investigate the nonlinear evolution of the instability in an idealized configuration, ruling out the development of secondary Rayleigh–Taylor instabilities.[39–41]

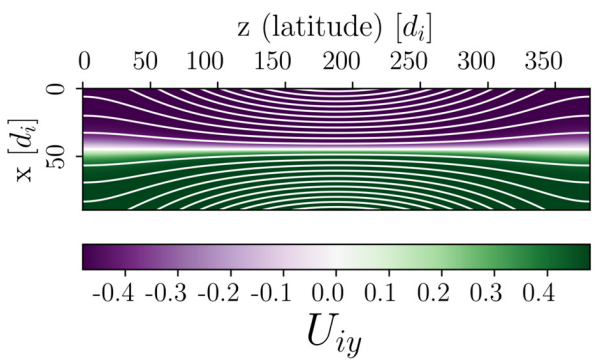

FIG. 2. Isocontours of the magnetic flux function $\psi(x, z)$ (white solid lines) and the spatial distribution of the ion velocity $U_{iy}(x, z)$ at equilibrium $(t = 0)$. At the center of the box along the $x$-axis, there is a white band indicating the frontier between the MS and the SW plasmas at equilibrium.

We must underline that both runs A and B are very idealized. They can be seen as the two "reference cases" for studying MLR reconnection only and type I VIR. The advantage of taking a symmetric magnetic field in run B (with a flow-aligned component going from $-\tan\phi/2$ to $+\tan\phi/2$ as in Refs. 5 and 11) is that the most unstable mode of the KHI develops with the same wavevector that develops in run A (same $y$-direction and same wavelength), allowing for a one-to-one comparison between the two runs. We can, thus, inquire the role of the KHI and high-latitude stabilization on the development of reconnection when only one parameter is changing (the initial current at the magnetopause). For the parameters we adopt, the FGM at the equator has $m = 2$, where $m$ is the mode number along the $y$-direction. Except for the density, the chosen parameters (e.g., the ion skin depth $d_i$, Alfvén velocity $v_A$, and the other normalization quantities) represent typical values of physical quantities observed in the outer MS, with magnetosonic and Alfvénic Mach numbers of order $M_f \simeq 0.8$ and $M_A \simeq 0.9$, respectively.

For practical reasons, we define a scalar quantity $\zeta$ that is passively advected by the plasma velocity, following $(\partial/\partial t + \mathbf{u} \cdot \nabla)\zeta = 0$, with $\zeta = 0.6 + 0.4 \tanh(\psi/L_{eq})$. This passive tracer helps us visualize plasma structures even in the absence of a density jump across the magnetopause. Indeed, the MS plasma is characterized by values $\zeta < 0.6$, while the SW plasma has $\zeta > 0.6$, with the magnetopause surface corresponding to $\zeta_{fr} = 0.6$.

## III. RESULTS AND SIMULATIONS COMPARISON
### A. Asymmetrical vs symmetrical development of vortices and reconnection events

In Fig. 3, we show a 3D rendering of the nonlinear evolution of runs A and B (panels A1 and B1, respectively). The 3D boxes in this and the following figures have been doubled along the $y$ periodic direction ($L_y \to 2L_y$) for the sake of clarity. First, we note that high latitudes remain stable, as shown by the lack of structures at the planes $z = 0, L_z$, while folded vortices have formed around the equatorial region in both simulations. Second, while in run A vortices develop almost symmetrically with respect to the equatorial plane, in particular, they develop at around $z = L_z/2$, run B exhibits an asymmetric evolution with more developed vortices in the southern hemisphere, in particular, at around $z \simeq 115 d_i < L_z/2$. Data from run A and run B are plotted at different times, $t = 460 \,\Omega_c^{-1}$ and $t = 500 \,\Omega_c^{-1}$ respectively, but corresponding to the same phase of the instability's evolution. This apparent mismatch between the instability stage and the simulation time is due to the different growth rates for the two runs because of the stabilizing effect of the magnetic shear on the KHI in run B. To compute the growth rate of the instability at the equatorial plane, we can approximate our system to a slab configuration of two uniform inviscid magnetized fluids in relative motion. The dispersion relation[42,43] suggests that the growth $\gamma_B$ for run B should be slightly smaller than $\gamma_A$, the one for run A. Indeed, during the early development of the KHI, we infer from simulation data $\gamma_A = 0.0112 \,\Omega_c \,(= 0.0672 \,\text{s}^{-1})$ and $\gamma_B = 0.0088 \,\Omega_c \,(= 0.0528 \,\text{s}^{-1})$. Given that the folding time of the vortices is inversely proportional to the growth rate of the KHI,[44] run B takes longer to reach the folding phase of the instability.

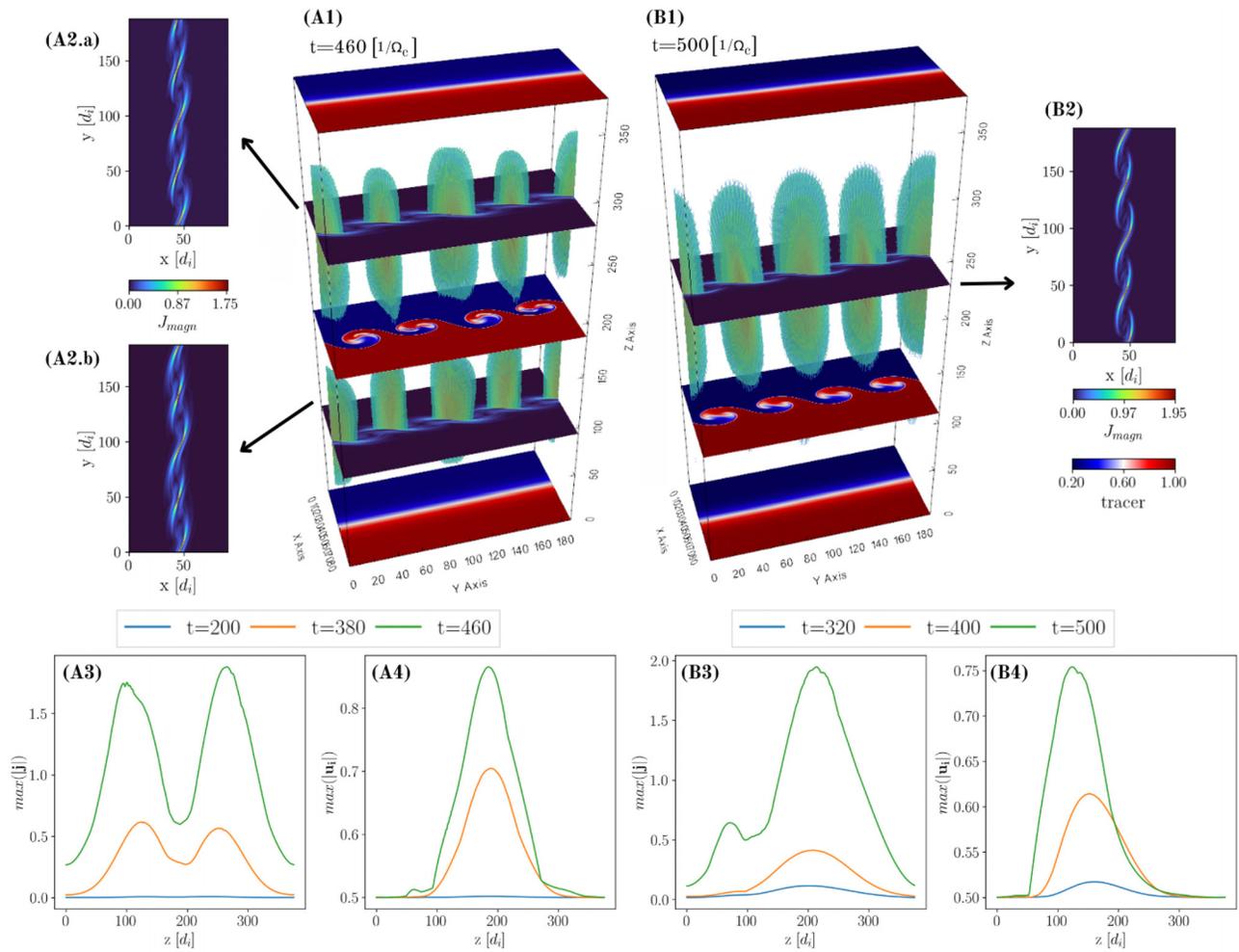

**FIG. 3.** Panel (A1): 3D representation of the current sheets and KH vortices forming at $t = 460\Omega_c^{-1}$ for run A. Slices at latitudes $z = L_z$, $z = L_z/2$, and $z = 0$ show the latitudinal evolution of vortices via the passive tracer [red = SW and blue = MS, as shown in the colorbar below panel (B2)]. Green current sheets form around mid-latitudes. Panel (A2.a-b): Two slices at $z = 3L_z/4$ (A2.a) and $z = L_z/4$ (A2.b) show the magnitude of the current density $|\mathbf{j}|$. Panel (B1): 3D representation of the current sheets and KH vortices forming at $t = 500\Omega_c^{-1}$ for run B. Slices at latitudes $z = L_z$, $z = 115d_i$, and $z = 0$ depict the latitudinal evolution of vortices via the passive tracer [red = SW and blue = MS, as shown in the colorbar below panel (B2)]. Green current sheets form primarily in the northern hemisphere. Panel (B2): Slice at $z = 240d_i$ showing the magnitude of the current density $|\mathbf{j}|$. Panels (A3) and (A4): Latitudinal distribution of the maximum of the current density $\max(|\mathbf{j}|)$ and the ion velocity $\max(|\mathbf{u}_i|)$, respectively, at $t = 200, 380, 460\Omega_c^{-1}$ for run A. Panels (B3) and (B4): Latitudinal distribution of the maximum of the current density $\max(|\mathbf{j}|)$ and the ion velocity $\max(|\mathbf{u}_i|)$, respectively, at $t = 320, 400, 500\Omega_c^{-1}$ for run B.

The symmetric and asymmetric evolution in the two runs is also visible in the development of the current sheets. In the symmetric case, the formation of current sheets is driven by the sole differential advection of magnetic field lines.[20] In run A, we show that magnetic field lines anchored at high latitudes in the unperturbed plasma undergo counterstreaming flows and, simultaneously, are advected at low latitudes by the growing KH vortices. Since vortices are nearly at rest in our frame, lines belonging to different plasma regions are bent in opposite directions. This process results in the mostly symmetric formation of current sheets, highlighted as green sheets with peaks (reddish regions) near mid-latitudes in panel (A1). At those latitudes ($z = L_z/4, 3L_z/4$), the slices in panel (A2.a and A2.b) display the magnitude of the current density, showing the current sheets between vortices.

Panels (B1-2) of Fig. 3 refer to run B at $t = 500\Omega_c^{-1}$. In panel (B1), current sheets lie primarily in the northern hemisphere, with the green highlighted region expanding across the upper part of the numerical box asymmetrically with respect to the equatorial plane. The peaks in the current sheets (reddish regions) are located around $z = 240d_i > L_z/2$, where the horizontal slice in panel (B2) displays the magnitude of the current density. In this case, current formation is due to a combined action. Current pinching, imposed by the growing KHI, enhances the original current intensity. At the same time, differential advection modifies the original magnetic rotation by arching magnetic field lines, thus increasing the current in the northern hemisphere while decreasing rotation and current in the southern one. The hemisphere where the current is enhanced is determined by the scalar product between the equilibrium current and the equilibrium vorticity:[45] $\boldsymbol{\omega}_{eq} \cdot \boldsymbol{j}_{eq}$, where $\boldsymbol{\omega}_{eq} = \nabla \times \boldsymbol{u}_{eq}$ and $\boldsymbol{j}_{eq} = \nabla \times \boldsymbol{B}_{eq}$ are the equilibrium vorticity and current, respectively.

The symmetrical development of current sheets and vortices with respect to low latitudes for Run A is further displayed in Fig. 3 panels (A3) and (A4). The latitudinal distribution of the current density magnitude in panel (A3) exhibits two peaks, which develop with the beginning of the nonlinear phase of the KHI and grow as the simulation proceeds. Instead, the latitudinal distribution of the ion velocity in panel (A4) has a single peak around the equator ($z = L_z/2$), where the KH vortices fully develop. On the other hand, the simulation with an initial magnetic shear displays an asymmetric evolution with respect to the equatorial plane, as shown in Fig. 3 panels (B3) and (B4). The latitudinal distribution of the magnitude of the current density (B3) starts with a single peak at $t = 320\Omega_c^{-1}$ due to the pinching of the equilibrium current sheet induced by the velocity field imposed by the growing KHI. This peak then shifts toward the northern hemisphere ($z > L_z/2$) since differential advection increases the magnetic rotation there. A second, smaller peak develops at later times ($t = 500\Omega_c^{-1}$) in the southern hemisphere at $z \simeq 80 d_i < L_z/2$. Indeed, after differential advection has lowered the magnetic angle $\phi_B$, the latter changes sign and creates local current sheets with $\boldsymbol{j}$ oppositely directed with respect to the equilibrium current. Considering the evolution of KH vortices, the ion velocity latitudinal distribution in panel (B4) shows a single peak that moves toward the southern hemisphere. Consequently, there is a southward latitudinal shift of the vortices as the KHI is inhibited in the northern hemisphere. Indeed, the larger the magnetic rotation ($\phi_B$ increases in the northern hemisphere), the larger the impact of the magnetic tension on the KHI.[42,43]

As the development of the KHI dominates the plasma dynamics in both simulations, large-scale vortex structures form, as shown in Fig. 4. The passive tracer enables us to follow the morphological evolution of the vortices for run A (top row) and run B (bottom row). For both configurations, we show the different phases of the evolution: several vortices fully form (left column) and start pairing (middle column) but get disrupted by secondary KH instabilities before the full pairing (right column). In Fig. 3, we have shown that vortices are enhanced at different latitudes, depending on the initial magnetic rotation. For this reason, in Fig. 4, the horizontal slices ($(x, y)$-planes) are taken at $z = L_z/2 \simeq 188 d_i$ for run A and at $z \simeq 105 d_i$ for run B, i.e., at the equator and slightly below it, respectively.

### B. Quantitative analysis of reconnecting field lines

To obtain quantitative insight into the reconnection process brought about by KH mixing, we measure the number of reconnected lines that are generated in both configurations by tracing $\sim 22\,000$ magnetic field lines in both runs at each time of the simulations. We start by identifying a region in the $(x, y)$-plane at $z = 0$ of size $[-L_x/6, L_x/6] \times [0, L_y]$ and proceed by selecting the original position of each magnetic field line evenly across this region, integrating the lines along the whole box.[46] At $t = 0$, being the equilibrium field lines tangent to $\psi$-isosurfaces, and thus, to $\zeta$-isosurfaces too, the passive tracer $\zeta$ is constant along each line. In an ideal evolution for the magnetic field, even if field lines are stretched and distorted, $\zeta$ would remain constant along each line, i.e., each line would maintain its original connectivity to the sole blue/red (in terms of the passive tracer color scale) plasma. This is what we observe during the early evolution of our simulations. As soon as the non-ideal terms in Ohm's law start to increase in importance, allowing for reconnection, a variation in $\zeta$ along a field line would underline the fact that this line has changed connectivity. For this reason, we compute the value of the passive tracer along each line, which enables us to determine the latitude at which a given line crosses the magnetopause and changes its connection and to keep track of how many times a line undergoes such a process. In order to ensure that a line is crossing the frontier between the two plasma regions, we select a threshold of $\zeta_{jump} = 0.05$, and we define a line as "reconnected" if it exhibits a variation of $\zeta$ of, at least, $2\zeta_{jump}$ across the magnetopause. In this way, if, for example, a line originates at the bottom of the numerical box at $\zeta_{in} < \zeta_{fr} - \zeta_{jump}$ (i.e., in the MS region), then it is counted as reconnected only if the value of the passive tracer computed along the line itself, as it goes up the box, reaches anywhere a value $\zeta > \zeta_{fr} + \zeta_{jump}$. Hence, we define a connection change (CC) as the crossing of the magnetopause (satisfying the threshold condition) by a reconnected line. For a given line, the number of CCs, from the bottom to the top of the box, defines it as a once-, twice-, or more than twice-reconnected.

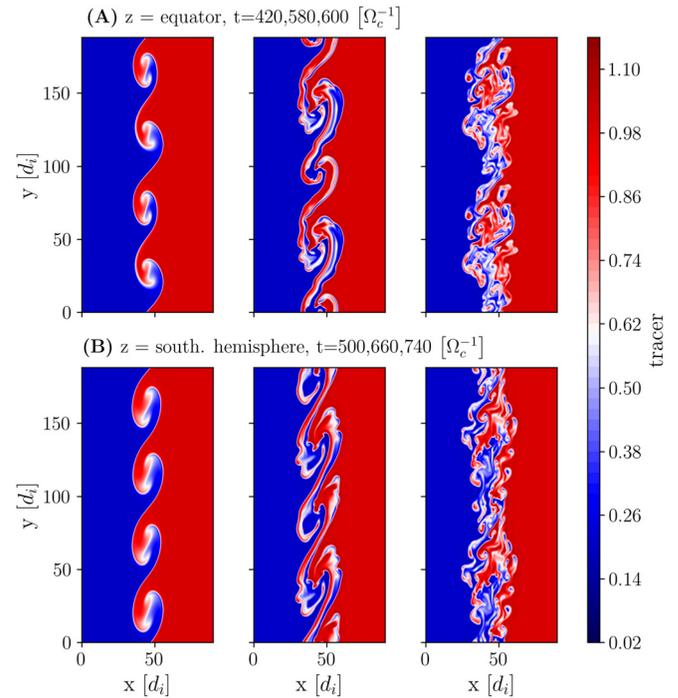

**FIG. 4.** Slices of run A at the equator (top row) and run B below the equator in the southern hemisphere (bottom row) showing the temporal evolution of the KH vortices through the passive tracer $\zeta$ that highlights the magnetosheath plasma (in red) and the magnetospheric plasma (in blue).

Figure 5 presents the cumulative count of CCs across both simulations, as a function of $z$, for three different times. Panel (A) shows how, in run A, the latitudinal distribution of CCs develops two peaks at mid-latitudes in both hemispheres, displaying quantitatively for the first time a nearly symmetrical evolution with respect to the equator $z = L_z$. This pattern aligns with the results depicted in Fig. 3 panel (A3), where current sheets manifest at mid-latitudes. Consequently, the development of current sheets at mid-latitudes initiates magnetic reconnection, resulting in a concentration of CCs for the analyzed field lines, precisely at the latitudes where the earlier current sheets formed. This configuration for the current sheets (at mid-latitudes, away from the vortices), combined with the distribution of reconnection events, suggests that only the double-MLR process is active in this simulation.

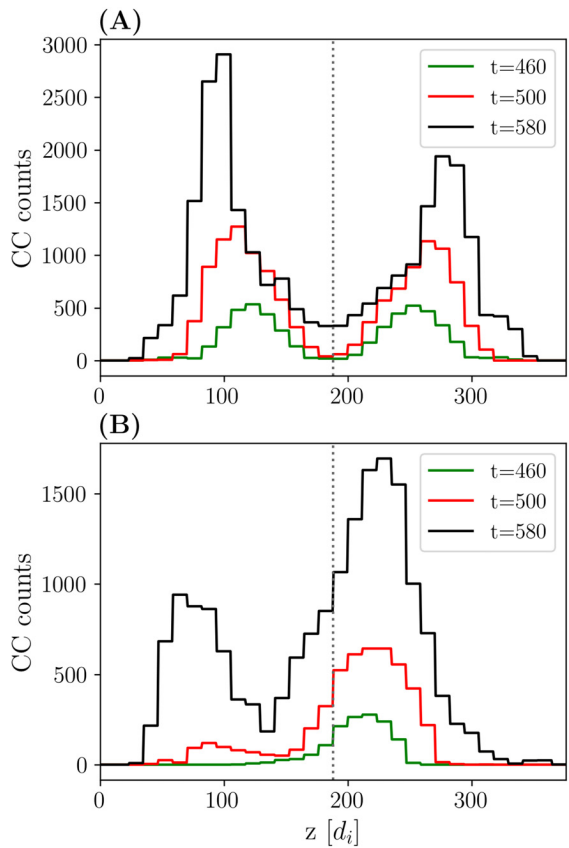

**FIG. 5.** Histograms of the CC counts (for $\zeta_{jump} = 0.05$) for both run A (top) and B (bottom) during the nonlinear phase of the instability at $t = 460$, 500, and $580\Omega_c^{-1}$. The gray dotted line in both panels indicates the z-position of the equatorial plane.

In Fig. 5(b), the profiles of the latitudinal distributions in Run B develop in time asymmetrically with respect to the equatorial plane. At $t = 460\Omega_c^{-1}$, there is only one main peak, which then migrates toward the northern hemisphere as time progresses. Already at $t = 500\Omega_c^{-1}$, a second, smaller peak forms in the southern hemisphere and proceeds to grow and shift toward the mid-latitudes of that hemisphere ($z \sim 80 d_i$). In this configuration, the position of the vortices with respect to the most prominent current sheets suggests that both MLR and type-I VIR are active. The main current sheet in the northern hemisphere (see Fig. 3 panel (B3)) is the original current sheet already present at the equilibrium that has shifted northward because of the differential advection of the field lines, becoming squeezed by the KH vortices, thus leading to the corresponding peak in the distribution of reconnection events (see Fig. 5(b)), compatibly with type-I VIR events. It is important to note that in 3D two different mechanisms are at work in modifying the current: pinching of the current sheet and differential advection. Nevertheless, since reconnection is occurring in the original current sheet, magnified by these mechanisms, and in particular, by the pinching, we call it type I VIR. Instead, the second, smaller peak in the distribution of reconnection events is consistent with MLR occurring in the smaller current sheets forming away from the equator. Interestingly, when comparing the two runs at similar stages of the KHI evolution, the sum of CCs is of the same order of magnitude. For example, at $t = 460\Omega_c^{-1}$ for run A and at $t = 500\Omega_c^{-1}$ for run B, the sum of CCs is of the order of $8 \times 10^4$ and $7.7 \times 10^4$, respectively for run A and B.

Figure 6 illustrates the temporal progression of CC counts for the reconnected magnetic field lines, showcasing the topological evolution during the nonlinear phase of both simulations. In particular, it shows how many lines have reconnected once (solid black line), twice (dashed black line), or more than twice (dotted black line) during the nonlinear phase. As a consequence of the different growth rates of the instability in the two runs, the nonlinear phase starts at slightly different times for the two simulations: around $t = 380\Omega_c^{-1}$ for run A and $t = 420\Omega_c^{-1}$ for run B. As previously mentioned, we can identify three primary stages in the vortex evolution for both runs: vortex formation, pairing, and disruption. These stages aid in interpreting the data in Fig. 6. Consequently, we define the start of the pairing phase at $t = 480\Omega_c^{-1}$ and $t = 580\Omega_c^{-1}$ for runs A and B, respectively. The disruption phase, instead, starts around $t = 580\Omega_c^{-1}$ and $t = 660\Omega_c^{-1}$ for runs A and B, respectively.

In Fig. 6(a), for a short initial period, the percentage of twice-reconnected lines in Run A is of the same order as the percentage of once-reconnected lines. However, when the vortices are fully formed (at $t = 460\Omega_c^{-1}$), right before they start pairing, we observe that the percentage of twice-reconnected field lines becomes dominant. This evolution that sees a dominance in twice-reconnected lines also continues during the pairing phase and is compatible with what is expected from a symmetric evolution of current sheets, which develop at mid-latitudes as shown in Figs. 3(A1) and 3(A2.a-b). Thus, this information on the twice-reconnected lines further supports the idea that only double-MLR is active when an initial magnetic rotation is absent. In Fig. 6(b), instead, the percentage of once-reconnected lines in run B is dominant throughout the evolution until the start of the disruption stage, where at $t \sim 660\Omega_c^{-1}$ we observe a sudden drop and the percentage of twice- and more than twice-reconnected lines becomes more important. The dominance of only once-reconnected field lines is also in accordance with what was earlier discussed on the formation of main current sheets in the northern hemisphere, with the consequent dominance of type-I VIR. As the vortex disruption phase starts ($t = 580\Omega_c^{-1}$ in run A and $t = 660\Omega_c^{-1}$ in run B), the number of field lines that have reconnected more than twice gradually increases, overtakes the number of once-reconnected lines, and finally approaches the number of twice-reconnected lines. The increase in reconnection events along each line can be justified by the start of secondary instabilities, like the development of secondary KHIs. In Fig. 7, we show indeed that for both runs, the large vortex structures are unstable to secondary KHI, again at different times because of the different growth rates of the primary instability. Small-scale KH vortices form along the arms of the primary vortices at latitudes at which they are most enhanced, specifically the equatorial plane in run A and the southern hemisphere in run B. Since the current is non-zero, the formation of secondary vortices helps to reduce the thickness of current sheets and, thus, facilitate magnetic reconnection. Since the current is not zero, the secondary vortices pinch it and boost the production of reconnected lines.[28] In particular, in Figs. 6(a) and 6(b), the dotted line, which represents the percentage of lines that reconnect many times, displays a sudden increase around the times at which the secondary KHI develops (see Fig. 7). In run A, the difference in percentage between the once- and twice-reconnected lines remains mostly constant at $\sim 20\% - 25\%$ from the start of the pairing phase ($t = 480\Omega_c^{-1}$) until the end of the simulation. On the contrary, the number of more than twice-reconnected lines suddenly increases as secondary KHI starts. In run B instead, this difference between once- and twice-reconnected lines (with, initially, once-reconnected lines dominating the dynamics) is not only non-constant but is even inverted. We see indeed that as the pairing stage starts ($t = 580\Omega_c^{-1}$), this difference is about $\sim 20\%$, but it decreases until there is an inversion and twice-reconnected lines become dominant (and far from negligible) at $t = 680\Omega_c^{-1}$, after the start of the disruption phase of the vortices. While the symmetric nature of run A naturally leads to (and confirms) a higher number of twice-reconnected lines compared to once- and more than twice-reconnected lines, it is noteworthy that their overall count is approximately the same as in run B at the end of the simulations.

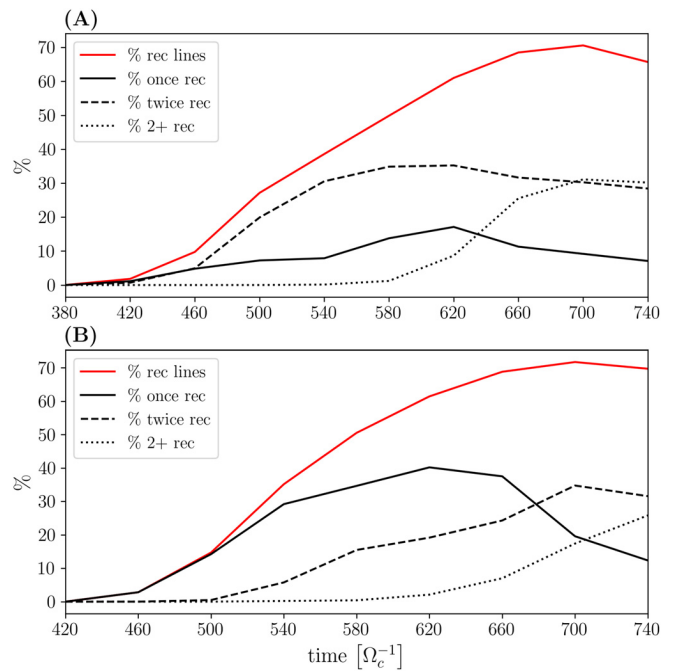

**FIG. 6.** The temporal evolution represents the count of CC as a percentage of the $\sim$22 000 integrated magnetic field lines at each time step. Both in (a) and (b), we show the percentage of magnetic field lines that reconnected once (black solid line), twice (dashed line), and more than twice (dotted line). The cumulative percentage is shown in red.

Magnetic reconnection is expected to lead to the development of a mixing layer across the magnetopause, in which the mixing between the two different plasmas is dominated by the parallel streaming of particles along reconnected lines.[4,24] This plasma mixing, naturally present in kinetic simulation, is not described by our fluid model. Nevertheless, we can estimate the plasma mixing by looking at the magnetic field line mixing. For this purpose, we define a quantity that gives us the perpendicular displacement of magnetic field lines with respect to the position it would have during an ideal evolution. We recall that $\psi$ at $t = 0$ can be seen as a line label and, at the same time, as a measure of the distance of a field line from the magnetopause ($\psi = 0$ isosurface). If we let $\psi$ evolve ideally (following an ideal advection equation, as the passive tracer does), $\psi$ for $t > 0$ would no longer describe the magnetic field; however, it would still measure the distance from the perturbed magnetopause folded by the KH vortices. We, thus, define the quantity $\Delta\psi = \max(\psi) - \min(\psi)$ along one line, which estimates the perpendicular displacement of the line due to the non-ideal evolution, i.e., not taking into account the contribution of the displacement related to the ideal evolution.

In Fig. 8, we show the evolution in time of the average squared displacement $\langle \Delta\psi^2 \rangle$, finding similar profiles for the two simulations.

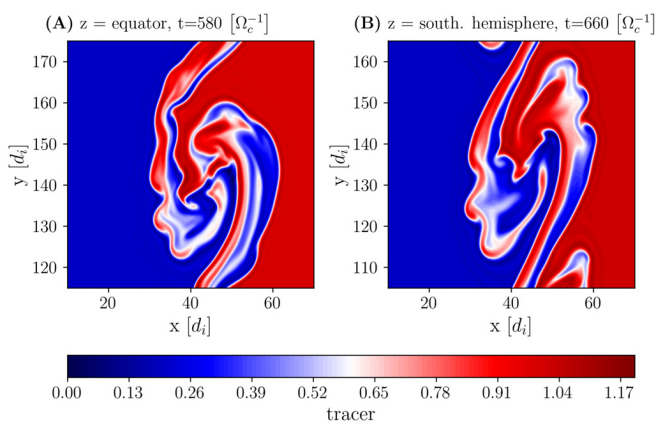

**FIG. 7.** Zoom-in onto a region of size $60 \times 60 d_i^2$ in the $(x, y)$-plane, highlighting secondary KHI developing in both runs. Left: run A at time $t = 580\Omega_c^{-1}$ in the equatorial plane ($z \simeq 188 d_i$). Right: run B at time $t = 660\Omega_c^{-1}$ in the southern hemisphere ($z = 105 d_i$).

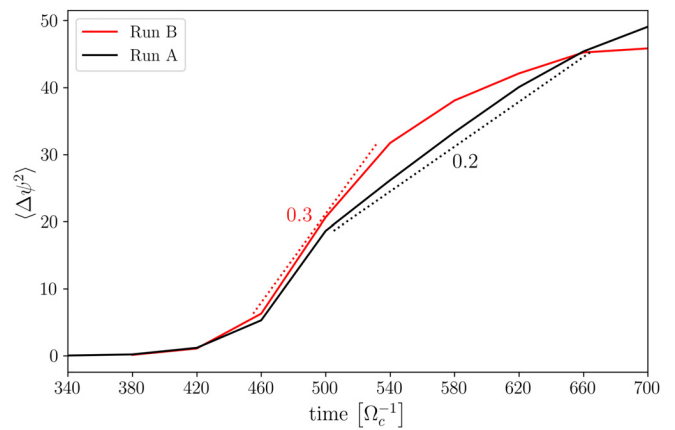

**FIG. 8.** Plot of the average displacement of magnetic field lines $\langle \Delta\psi^2 \rangle$ during the nonlinear phase with respect to their initial position for both simulations (solid red—run B and solid black—run A). Intervals of nearly linear growth (and their slope) are marked with dotted lines. Since the two simulations evolve at different rates and enter the nonlinear phase at different times, we shifted the profile of run B to match that of run A and ease the comparison.

This suggests that the presence of an initial magnetic shear does not significantly influence the movement of magnetic field lines during the actual system evolution with respect to the positions they would have in an ideal evolution. However, the profiles are not entirely overlapping, as we show in Fig. 8: run B reaches a plateau by the end of the simulation, while run A keeps increasing and does not appear to be stabilizing around a maximum value for the average displacement before the end of the simulation time. In both runs, we can identify intervals in time in which $\langle \Delta \psi^2 \rangle$ has a nearly linear growth (Fig. 8 dotted lines), indicating a diffusive-type widening of the mixing layer. In Run A, this interval lies between $t = 500 \Omega_c^{-1}$ and $t = 660 \Omega_c^{-1}$ with a slope of order 0.2, while in run B the slope of the linear growth is between $t = 460 \Omega_c^{-1}$ and $t = 540 \Omega_c^{-1}$ and is of order 0.3. The suggested effective magnetic and matter diffusion coefficient is, therefore, of order $D_{\text{eff}} \simeq q v_{A,z} d_i$, with $q$ the slopes mentioned above, and $v_{A,z}$ the z-component of the Alfvén velocity. Taking $v_{A,z} \simeq 400 \text{ km s}^{-1}$ and $d_i \simeq 250 \text{ km}^{28}$ we obtain $D_{\text{eff}}(\text{Run A}) \simeq 2 \times 10^{10} \text{ m}^2 \text{ s}^{-1}$ and $D_{\text{eff}}(\text{Run B}) \simeq 3 \times 10^{10} \text{ m}^2 \text{ s}^{-1}$.

## IV. DISCUSSION & SUMMARY OF THE RESULTS

In this section, we discuss and summarize the main results we obtained from the direct comparison between the two simulations and with previous works in the literature as follows:

(1) In both simulations, the KHI dominates the plasma dynamics, with the creation of large-scale vortices at low-latitude (folding phase) that merge (pairing phase) and finally get disrupted by secondary processes (disruption phase). The dispersion relation[42,43] foretells that run B should take longer to reach the folding phase of the instability. Indeed, during the early development of the KHI, we infer from data the growth rates $\gamma_B = 0.0088 < \gamma_A = 0.0112$. As a consequence, run B reaches each phase of the instability at a later time than run A.

(2) Focusing on the latitudinal evolution of KH vortices, they initially grow at low latitudes, but for a configuration with a purely northward magnetic field (run A), the system exhibits a nearly symmetrical development of the plasma dynamics with respect to the equatorial region, until the end of the simulation. On the contrary, in run B, vortices quickly migrate toward the southern hemisphere. This is due to the differential advection of magnetic field lines, tied at high latitudes to the unperturbed plasma but advected at low latitudes by the KH dynamics. Differential advection decreases the magnetic shear angle $\phi_B$ in the southern hemisphere, enhancing the KHI there, while it increases $\phi_B$ in the northern hemisphere, weakening the instability's growth there, with the consequent migration of vortices. This behavior is in agreement with previous works;[28,29,45] in particular, the asymmetrical development of the system observed in run B is determined by the sign of $\boldsymbol{\omega}_{\text{eq}} \cdot \boldsymbol{j}_{\text{eq}}$. In our case, $\boldsymbol{\omega}_{\text{eq}} \cdot \boldsymbol{j}_{\text{eq}} > 0$, so vortices develop in the southern hemisphere, as expected. Contrary to the previous works, because of the symmetry in the rotation of the initial magnetic field in run B, we do not introduce a tilt in the vortex axis. This choice allows for a one-to-one comparison between the two different configurations (with and without the magnetic shear), having both the vortex axes aligned in the z-direction.

(3) Our analysis of the current sheet development shows a similar evolution, with a symmetrical formation of current sheets in run A and an asymmetrical one in run B with respect to the equatorial plane. The phenomenon of differential advection induces a bending of magnetic field lines, generating current sheets. In run A, with zero initial current, current sheets form at mid-latitudes. Conversely, in run B, where an initial current was present, the current is enhanced in the northern hemisphere by both current pinching and differential advection. Only when the system is well into the nonlinear phase does a second smaller current sheet develop in the southern hemisphere below the latitude at which vortices form. Indeed, differential advection initially lowers $\phi_B$ in the southern hemisphere, reducing the current, but eventually, it is able to reverse $\phi_B$, creating new current sheets there.

(4) The latitudinal distribution of the connection changes (CCs) shows a qualitatively similar symmetrical/asymmetrical evolution for the two simulations, with a clear correspondence between the peaks of the latitudinal distributions of CC count and of the current density magnitude. In run A, the configuration of current sheets at mid-latitudes, away from the KH vortices, combined with the latitudinal distribution of reconnection events, shows that only double-MLR is active. This was suggested in earlier works,[20,46] but the CC count was never quantified. In run B, the vortex positions relative to the main current sheets suggest simultaneous MLR and type-I VIR activity. The main current sheet, which drifted toward the northern hemisphere, causes a peak in reconnection events associated with type-I VIR. Meanwhile, a second smaller peak in reconnection events is consistent with MLR, occurring in less intense current sheets away from the equator, in the southern hemisphere. The similar characteristics of the two simulations enabled us to quantitatively compare the number of reconnection events at similar times during the systems' evolution. Our analysis shows that the count of all reconnection events along the computational box for each run is of the same order of magnitude: $\sim 8 \times 10^4$ at the folding phase, $\sim 2 \times 10^5$ at the merging phase, and $\sim 4 \times 10^5$ when the onset of secondary KHI disrupts the large-scale vortex structures.

(5) Because of the design of the two simulations, identical in size and very similar in initialization, for the first time, we can study and compare quantitatively the topological evolution of magnetic field lines in these systems. During the nonlinear phase, we found that, in run A, there is a clear dominance of lines that reconnect twice. The number of twice-reconnected lines is constantly $\sim$ twice that of once-reconnected lines from the start of the pairing phase until the end of the simulation time for run A. Instead, in run B, once-reconnected lines dominate during the initial evolution of the system, but their dominance ends when secondary processes disrupt the KH vortices. It is essential to underline that even if once-reconnected lines are dominant in run B, the number of twice-reconnected lines is far from insignificant. For the first time, we highlight that by the end of both simulations, the proportion of twice-reconnected lines comprises nearly identical percentages, approximately 43% and 45%, of the total number of reconnected lines in runs A and B, respectively. Twice-reconnected lines, indicating multiple reconnection events and mixing of material, play an important role in determining not only the plasma transport across the magnetopause but also in allowing the specific entropy to increase when passing from the magnetosheath to the magnetosphere.[21,23]

(6) In earlier 2D[39–41] and 3D[5,28,29] works, it was observed that, after the folding phase, the conditions were favorable for the development of secondary KH and Rayleigh-Taylor instabilities inside the vortices. In all these cases, the Atwood number of the equilibrium configuration was different from zero. A density jump between the plasmas on the two sides of the magnetopause is necessary for the development of secondary Rayleigh-Taylor instability, for which the vortex rotation provides the effective gravity. This difference is also needed, at least in 2D, to have strong vorticity layers along the vortex arms and, thus, secondary KHI. We observe that, even in the absence of a density jump in the initial setting of both simulations, we see the formation of secondary KH vortices along the arms of primary KH vortices. These dynamics suggest that the vortex layers where secondary KH vortices grow are created by 3D effects, in particular by the combined action of magnetic tension and line tying. These secondary processes induce magnetic reconnection at intermediate latitudes between the two peaks in the CC count. Thus, they influence the number of times each magnetic field line reconnects and increase the amount of line reconnected more than twice, which reaches the total percentage of ∼46% and ∼36% in run A and run B, respectively.

(7) Finally, the effective magnetic diffusion coefficient $D_{\text{eff}}$ associated with reconnection is of order $\sim 10^{10}$ m$^2$s$^{-1}$ for both simulations. In the literature, there is no general consensus concerning the quantification of this coefficient. A lower limit to explain observed plasma mixing is given in a simplified diffusive model of the magnetopause[47] and is of $\sim 10^9$ m$^2$ s$^{-1}$, which was later confirmed by simulations.[5,28] Another 3D MHD work[48] finds the same order of magnitude as our work, but 3D kinetic simulations[11] found instead an effective diffusion coefficient one order of magnitude greater: $\sim 10^{11}$ m$^2$ s$^{-1}$.

## V. CONCLUSIONS

In this paper, we presented our work on 3D resistive Hall-MHD numerical simulations that reproduce the KHI dynamics at the Earth's magnetospheric flanks. Our focus was on the interaction of SW and Earth's MS, driven by the development of KHI and induced magnetic reconnection events. We have compared for the first time two identical 3D configurations, including high-latitude stabilization, only differing by the presence or absence of a large-scale magnetic shear. Even if in idealized configurations, our 3D framework captures three-dimensional processes, such as high-latitude stabilization of KHI and double ML reconnection events.

The design of these simulations allows for a direct comparison between the two scenarios of KHI development, highlighting the role of the nonlinear mechanisms (differential advection in both runs A and B and current pinching in run B) in determining the differences in the KH vortex evolution (symmetric/asymmetric evolution) and in the induced reconnection dynamics.

For the first time, we have quantitatively analyzed once, twice, and more than twice reconnected magnetic field lines in a configuration in which only ML reconnection is active. We have then performed the same analysis on the other run, allowing for a quantitative comparison in the two runs of plasma dynamics and magnetic reconnection events.

In summary, our comparative analysis of these two simulations stresses the crucial role of magnetic field orientation in shaping the Earth's flank magnetopause. Specifically, it influences vortex migration, current sheet formation, and the distribution of reconnection events in latitude. These findings highlight the necessity of considering 3D effects and line tying for an accurate portrayal of the complex phenomena ruling the flank dynamics.

## ACKNOWLEDGMENTS

F.B. acknowledges the support from the FED-tWIN programme (profile Prf-2020-004, project "ENERGY") issued by BELSPO and from the FWO Junior Research Project No. G020224N granted by the Research Foundation – Flanders (FWO). This work was granted access to the HPC resources of IDRIS under the allocation 2022-AD010413669 made by GENCI.

## AUTHOR DECLARATIONS
### Conflict of Interest

The authors have no conflicts to disclose.

### Author Contributions

**Silvia Ferro:** Data curation (lead); Formal analysis (lead); Investigation (lead); Validation (lead); Visualization (lead); Writing – original draft (lead). **Matteo Faganello:** Conceptualization (lead); Funding acquisition (lead); Methodology (lead); Project administration (lead); Resources (lead); Software (lead); Supervision (lead); Writing – review & editing (equal). **Francesco Califano:** Conceptualization (lead); Funding acquisition (lead); Investigation (supporting); Supervision (supporting); Writing – review & editing (supporting). **Fabio Bacchini:** Funding acquisition (supporting); Project administration (supporting); Resources (supporting); Supervision (supporting); Writing – review & editing (equal).

## DATA AVAILABILITY

The data that support the findings of this study are available from the corresponding author upon reasonable request.